# Study of number of particles crossing through a scintillation detector

Mahmoud reza Oshagh[1], Shoubane Hemmati[2], Farnaz Behrouzi[2],
Farzaneh Sheidaei[2], Mahmoud Bahmanabadi[2]

[1]Department of physics, Sharif university of technology, Iran.
[2]Department of physics, Tarbiat Modares University, Iran

*Abstract*. **In this study we set our system in order to study the energy spectrum of single, double and triple particles, detected in a scintillation detector. The goal of doing this experiment was to determine the probability of number of particles (single, double or triple) detected, from the energy spectrum in any given energy spectrum. The results of experiment will be used in our extensive air shower array.**

I. INTRODUCTION

In the typical extensive air shower (EAS) experiment, shower particles produced in the atmosphere by ultra-high energy($>10^{13} eV$) gamma-rays or cosmic rays(CR) are detected by a distributed array of large area plastic scintillation detectors [1] and some times other types of particle detectors for improvement of accuracy. The arrival direction of an air shower can be determined from fast timing data of the detectors and the accuracy of the obtained direction depends on the accuracy of time measurement. The detectors can also determine the local density of the shower particles and the position of the core of EAS [2]. The purpose of this study is to determine the probability of number of particles (single, double or triple) detected, from their deposit energy spectrums.

II. EXPERIMENTAL SETUP

*A. single particle detection*

In this study we set our system in order to find the deposit energy spectrum of single particle, crossing through a scintillation detector. In our setup, we used a big plastic scintillator (NE102) with a size of $100 \times 10 \times 1 cm^3$ and on top of it we used a same plastic scintillator but with size of $10 \times 10 \times 1 cm^3$. Each one of them has a single photomultiplier tube (PMT, EMI 9813 KB) located at its $10 \times 1 cm^3$ side via a regular plastic light guide. This setup was arranged as shown in Fig.1, in cosmic ray Laboratory of Department of physics at Sharif University of Technology in Tehran. The signals produced by secondary particles of an extensive air shower (EAS) are entered to a time to amplitude converter (TAC) via an 8-fold fast discriminator. The output of the scintillator #1 is connected to the start input of TAC. The output of the scintillator #2 is connected to the stop input of TAC.

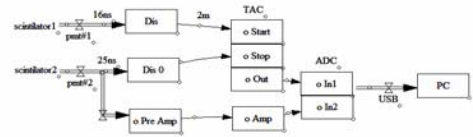

Figure 1: schematic view of experimental number 1

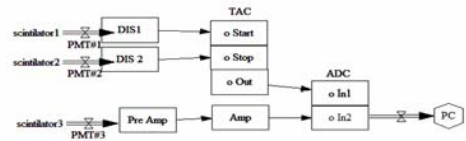

Figure 2: schematic view of experimental number 2

The dynode output of big scintillator is connected to a preamplifier that amplifies the voltage by 100 and output of preamplifier is connected to the input of amplifier that amplifies the voltage by 500. Then the output of TAC and the output of amplifier are fed into a multiparameter multichannel analyzer (MCA) via an analogue to digital converter (ADC) unit. The output of TAC triggers the ADC, and the time lag between the output signals of PMTs (#1 and #2), and the pulses of big scintillator passed from amplifier are read out as parameters 1 and 2 respectively. So by this procedure an event is logged. We put the small scintillator on different places of the big scintillator, so that covers the whole of surface of the big scintillator. Each data set covers a total period of 3600 seconds.

*B. double particle detection*

For detection double particle we used three scintillators containing two scintillators with size of $10 \times 10 \times 1 cm^3$ and one $100 \times 10 \times 1 cm^3$. We put the two small ones on top of the big one. Figure 2 shows the arrangement. In this experiment the main idea was deposit energy of two particles detected coincidently. In order to achieve this aim we ran this experiment for periods of 24 hours with moving successively the two small scintillators on top of the big scintillator, and each time recording time lag of the two small scintillators and deposit energy of double particles in the big scintillator. Nine experiments which each of them covers a same time with first experiment.

*C. Triple particle detection*

The main aim of this part of the experiment was to gain the deposit energy spectrum of 3 particles crossing through the scintillator detectors coincidently. It is also a good estimation of whether the last two parts of the experiment achieved a satisfactory result. The setup of this part can be seen in Fig. 3. Three scintillators (Nos. 2, 3, and 4) with effective dimensions of $10 \times 10 \times 1 cm^3$ are placed on top of the big scintillator (No.1, $100 \times 10 \times 1 cm^3$). The gate of ADC comes from output of the first TAC which means the system will be triggered only if a particle passes through the first and second scintillators. It is also assumed that any particle which passes through upper scintillators will eventually pass through the big one and with this assumption we read the deposit energy from the big scintillator. Other parameters, which were essential for obtaining the correct deposit energy spectrum were the three time lags between the output signals of PMTs (1,2), (1,3), and (1,4).

We ran this experiment for 48- hour period putting the three scintillators (small ones) on top of the big one and moving those (together) 10 cm by 10 cm over the big scintillator. We did 8 experiments with a similar time.

III. DATA ANALYSIS AND DISCUSSION

Figure 4 shows the single particle deposit energy spectrum for the total deposit energy spectrum for all of the big scintillator area. Figure 5 shows the double particle deposit energy spectrum for the total deposit energy spectrum for all of the big scintillator area. For triple particle we had put the second, third and forth scintillators at 0, 10 and 20 cm of the PMT of the first scintillator respectively, and collected data for 2 days. Obtaining the deposit energy spectrum of single particle from this data, we chose the deposit energy of those particles which were in signals of time lag spectrum of one of them only. Similarly to obtain the deposit energy spectrum of double particles we chose the deposit energy of particles when were only in two time lag spectrums. And by choosing the deposit energy of particles which were in all three time lag spectrums, we obtained total deposit energy spectrum of triple particles.

By normalizing all of the plots into plots that the area under the curves to be one, we have compared them in Fig.6 By dividing the count of each channel of energy for single, double, and triple by total count of single, double and triple, we will be able to find the probability of finding single, double, and triple particles from the channel of energy that it is detected. Fig.7 shows the detection probability of single, double, and triple particles.

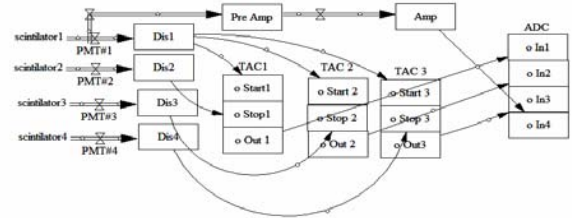

Figure 3: schematic view of experimental number 3

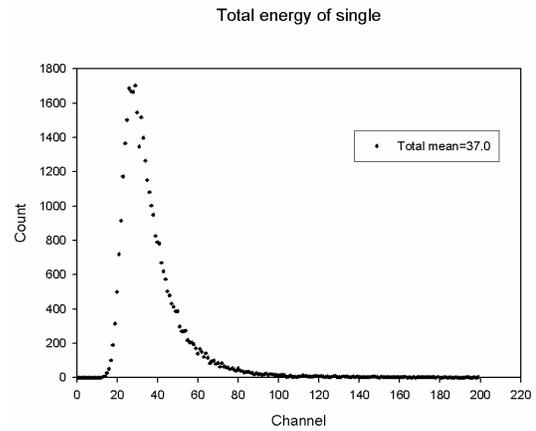

Fig. 4: energy spectrum of the total energy spectrum of single particle for experiment No.1.

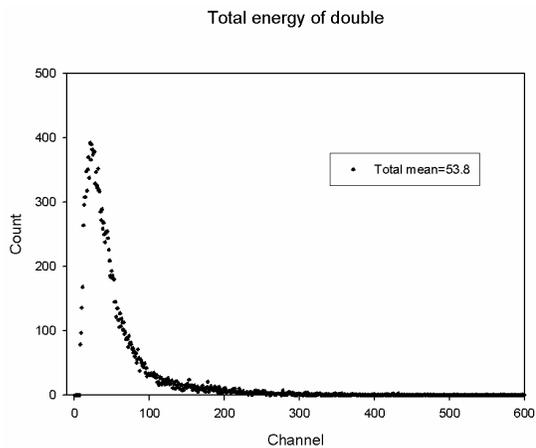

Fig. 5: energy spectrum of the total energy spectrum of double particles for experiment No.2.

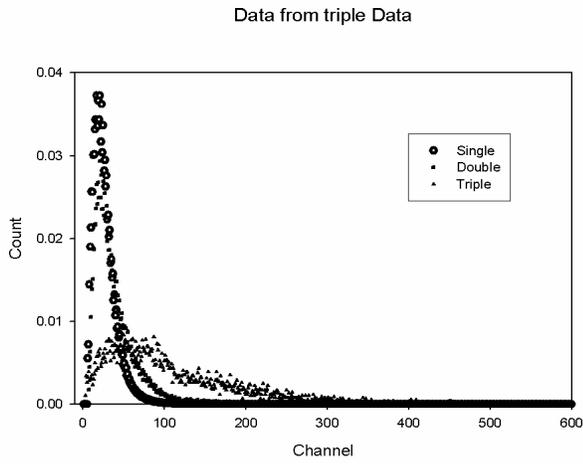

Figure 6. Normalized spectrum of energy of single and double and triple particles from exp no 3 data

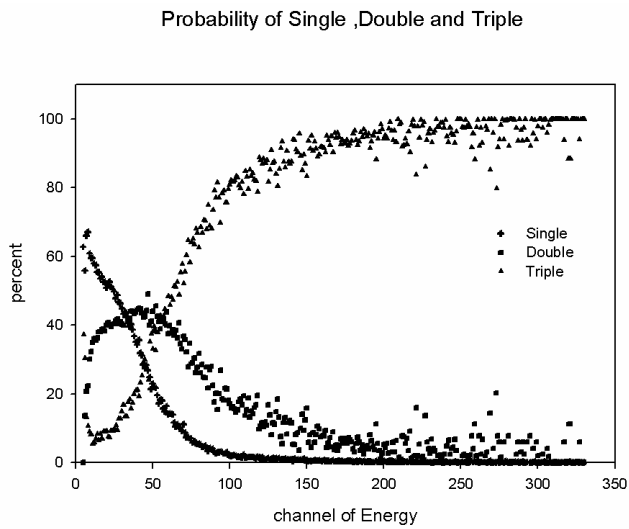

Figure7. Probability of finding single, double and triple particles